\documentstyle [12pt] {article}

\newcommand{\gtwid}{\mathrel{\raise.3ex\hbox{$>$\kern-.75em\lower1ex
\hbox{$\sim$}}}}
\newcommand{\ltwid}{\mathrel{\raise.3ex\hbox{$<$\kern-.75em\lower1ex
\hbox{$\sim$}}}}
\newcommand{\beq}{\begin{equation}}
\newcommand{\eeq}{\end{equation}}
\newcommand{\beqs}{\begin{eqnarray}}
\newcommand{\eeqs}{\end{eqnarray}}
\newcommand{\prl}{Phys. Rev. Lett.}
\newcommand{\prd}{Phys. Rev. D}

\catcode`@=11
\@addtoreset{equation}{section}
\@addtoreset{equation}{subsection}
\def\theequation{\ifnum\value{section}=0 \arabic{equation}\ignorespaces
\else \ifnum\value{section}=-1 A.\arabic{equation}\ignorespaces
\else \ifnum\value{subsection}=0 \thesection.\arabic{equation}\ignorespaces
\else \thesection.\arabic{subsection}.\arabic{equation}\ignorespaces
                           \fi
                      \fi
                 \fi}
\catcode`@=12

\begin{document}

\def\thefootnote{\fnsymbol{footnote}}
\baselineskip 7.5mm

\begin{flushright}
\begin{tabular}{l}
ITP-SB-93-68    \\
hep-ph/9403315  \\
March, 1994
\end{tabular}
\end{flushright}

\vspace{8mm}
\begin{center}
{\Large \bf Phases in Leptonic Mass Matrices:  }
\vspace{3mm}
{\Large \bf Higher Order Invariants and Applications to Models }

\vspace{16mm}

\setcounter{footnote}{0}
Alexander Kusenko\footnote{email: sasha@max.physics.sunysb.edu}
\setcounter{footnote}{6}
and Robert Shrock\footnote{email: shrock@max.physics.sunysb.edu}

\vspace{6mm}
Institute for Theoretical Physics  \\
State University of New York       \\
Stony Brook, N. Y. 11794-3840  \\

\vspace{20mm}

{\bf Abstract}
\end{center}

We discuss complex rephasing invariants of charged lepton and neutrino mass
matrices and associated theorems which determine in general (i) the number
of physically meaningful phases in these matrices and (ii) which elements
of these matrices can be rendered real by rephasings.  New results are
presented on higher order complex invariants and the application of the
methods to several models.

\vspace{35mm}
\pagestyle{empty}
\newpage

\pagestyle{plain}
\pagenumbering{arabic}
\renewcommand{\thefootnote}{\arabic{footnote}}
\setcounter{footnote}{0}

\section{ Introduction}
\label{intro}

   Understanding fermion masses and mixing remains one of
the most important outstanding problems in particle physics.  In particular,
the issue of possible neutrino masses and associated lepton mixing is of
fundamental interest.  Although there is no definite direct evidence for
nonzero neutrino masses\footnote{For reviews and current limits, see
Refs. \cite{pdg} and \cite{nurev}.  The apparent solar neutrino deficit is the
most suggestive indirect evidence at present. The situation with atmospheric
neutrinos is unclear.} they are expected on general grounds: given only the
known left-handed neutrino fields and the usual Higgs field(s) in the standard
model (and supersymmetric extensions thereof which stabilize the hierarchy),
nonzero neutrino masses result generically from higher-dimension operators
which are expected to occur at a scale near to that of quantum
gravity\footnote{This expectation is based on the general consensus that
pointlike theories of quantum gravity (in particular, supergravity) are
non-renormalizable and is borne out by the $E << (\alpha')^{-1/2}$ limit of
string theories.}, suppressed by associated inverse powers of the (reduced)
Planck mass, $\bar M_{P} = \sqrt{\hbar c/(8 \pi G_N)} = 2.4 \times 10^{18}$
GeV. For example, gauge-invariant dimension-5 operators of this type could
produce neutrino masses of order $m_\nu \sim a v^2/\bar M_P$, where
$v=250$ GeV is the scale of electroweak symmetry breaking, and $a$ is a
dimensionless constant.  (Here $\bar M_P$ is an approximate upper bound on the
mass which suppresses such operators; it is possible that new physics occurs at
some intermediate mass scale $v << M_I < \bar M_P$ in such a way that dimension
5 operators of this type would give rise to neutrino masses of order
$a' v^2/M_I$.) Thus one can understand on general grounds why
neutrino masses are so small.  It is not known whether there exist any
electroweak-singlet neutrino fields.  If they do exist, then
they could lead, via renormalizable, dimension-4 operators, to
neutrino masses $m_\nu \sim v^2/M_R$, where the scale $M_R$ of the
electroweak-singlet neutrino mass is naturally $>>v$, again yielding,
albeit for a different reason, very small $m_\nu$ \cite{seesaw}.

    In turn, a natural concomitant of
(nondegenerate) neutrino masses is lepton mixing, which is thus
also a generic expectation.\footnote{The lepton mixing angles are functions
of ratios of elements of neutrino matrix elements and of
charged lepton mass matrix elements, and even though left-handed neutrino
masses are small, some of these ratios could, in principle, be $O(1)$.
However, a set of conditions for natural suppression of observable lepton
flavor violation were formulated, and it was shown that the standard model
(generalized to include nonzero $m_\nu$) satisfies these \cite{bwlrs}.}
It is thus of interest to study models for the leptonic mass matrices in the
charge $Q=0$ and $Q=-1$ sectors.  The diagonalization of these matrices
determines both the masses and the observable lepton mixing.  In
addition to the search for neutrino masses and mixing in solar and
atmospheric neutrino experiments, searches continue at accelerators.  In
particular, the CHORUS and NOMAD experiments at CERN are currently
looking for $\nu_\mu \to \nu_\tau$ oscillations\footnote{The NOMAD
experiment uses a method proposed in Ref. \cite{as} while the CHORUS
experiment combines this with a search for the $\tau$ tracks in an
emulsion.}\cite{cern}

    In analyzing models of fermion masses and mixing, an important step
is to determine the number of real amplitudes and unremovable, and hence
physically meaningful, phases in the mass matrices.  Recently, we reported a
general solution to this problem \cite{ksl}.  Here we extend our explicit
construction of rephasing invariants and give further applications to various
models.  We also compare the situation concerning unremovable phases and
associated invariants in the leptonic sector with that in the quark sector, for
which we have also recently presented a general analysis \cite{ksq}.

The organization of this paper is as follows.
In section 2 we briefly review our theorem on the number of unremovable phases.
In section 3 we deal with the question of which elements of the mass matrices
can be made real by rephasings.  We discuss the connection between this and the
complex rephasing invariants.  This section includes explicit expressions
for the sixth order complex invariants.  In section 4 we apply
our general results to several models with and without electroweak singlet
neutrinos. Concluding comments are given in section 5.

\section{ Theorem on the Number of Physical Phases}
\label{number}

     The leptonic mass terms are taken to arise from interactions which are
invariant under the standard model gauge group $G_{SM}= SU(3) \times SU(2)
\times U(1)$, via the spontaneous symmetry breaking of $G_{SM}$.  In the
standard model and its supersymmetric extensions, the resultant mass terms
appear at the electroweak level via (renormalizable, dimension-4) Yukawa
couplings$^{5,6}$\addtocounter{footnote}{1}\footnotetext{We do not consider
models in which fermion masses
arise via multifermion operators, which are not perturbatively
renormalizable.  Note that at a nonperturbative level, lattice studies
\cite{yr} show that a lattice theory with a specific multifermion action
and no scalar fields may yield the same continuum limit as a theory with
a Yukawa interaction.}\addtocounter{footnote}{1}%
\footnotetext{In a number of interesting models, some of
these Yukawa couplings are viewed as originating at a higher mass scale,
such as that of a hypothetical supersymmetric grand unified theory (SUSY
GUT) or some other (supersymmetric) theory resulting from the
$E << (\alpha')^{-1/2}$ limit of a string theory.
It is possible that interactions which appear to
be Yukawa couplings at a given mass scale, could be effective, in the sense
that some of the elements of the associated Yukawa matrices could actually
arise from higher-dimension operators at a yet higher mass scale.
The plausibility of such
higher-dimension operators can be inferred either from the nonrenormalizability
of supergravity or as a consequence of the $E << (\alpha')^{-1/2}$
limit of a string theory.}. It follows that the mass terms for the charged
leptons can be written in terms of the $G_{SM}$ lepton fields as
\beq
-{\cal L}_{mass,\ell} = \sum_{j,k=1}^{3}\Bigl [(\bar L_{j L})_2
M_{jk}^{(\ell)}\ell_{k R}  \Bigr ] + h.c.
\label{ellmass}
\eeq
where $j,k=1,2,3$ denote generation indices, and
$L_{j L} = \Bigl (^{\nu_{_{\ell_j}}}_{\ell_j} \Bigr )_L$, are the left-handed
SU(2) doublets of leptons (where $\ell_1=e$, $\ell_2=\mu$, $\ell_3=\tau$).  The
index $a=2$ in $(\bar L_{jL})_2 = \bar \ell_{jL}$ is the SU(2) index.
Further, $\ell_{k,R}$ are the right-handed SU(2) singlets, and
$M^{(\ell)}$ is the charged lepton mass matrix. We use the result here from
LEP and SLC that there are three generations of standard model
fermion with associated light neutrinos \cite{pdg}.

In contrast to the charged lepton sector, where one at least knows the
relevant fields, in the neutral lepton sector, {\it a priori}, one does not;
in addition to the three known left-handed $I=1/2$, $I_3=1/2$ Weyl neutrino
fields $\nu_{jL}$, $j=1-3$, there could be some number $n_s$ of
electroweak-singlet neutrino fields $\chi_{j,R}$, $j=1,2,...n_s$.  The general
neutrino mass matrix is given by
\beq
-{\cal L}_{mass,N} =
 {1 \over 2}(\bar\nu_{_L} \ \bar{\chi^{c}}_{_L})
             \left( \begin{array}{cc}
              M^{(L)} & M^{(D)} \\
              (M^{(D)})^T & M^{(R)} \end{array} \right )\left( \matrix{
\nu^{c}_{_R} \cr \chi_{_R} \cr} \right ) + h.c.
\label{numass}
\eeq
where $\nu_{_L}=( \nu_e,\nu_\mu,\nu_\tau)_L^T$,
$\chi_{_R}=(\chi_1,...,\chi_{n_s})^T_R$; $M^{(L)}$ and $M^{(R)}$ are
$3 \times 3$ and $n_s \times n_s$ Majorana mass matrices, and $M^{(D)}$ is a
3-row by $n_s$-column
Dirac mass matrix.\footnote{In the context of supersymmetric extensions of the
standard model, we assume unbroken $R$ parity so that the neutrinos do not
mix with the neutralinos (higgsinos and neutral color-singlet gauginos).}
We denote the $(3+n_s) \times (3+n_s)$ neutrino mass matrix in
(\ref{numass}) as $M^{(N)}$, where $N$ denotes ``neutral'', $Q=0$.  (In Ref.
\cite{ksl} this was labelled equivalently as $M^{(0)}$.)
Thus,
\beq
M^{(L)}_{jk}=M^{(N)}_{jk}
\label{ml}
\eeq
where $j,k=1,2,3$;
\beq
M^{(R)}_{jk}=M^{(N)}_{j+3,k+3}
\label{mr}
\eeq
with $j,k=1,...n_s$; and
\beq
M^{(D)}_{jk}=M^{(N)}_{j,k+3}
\label{md}
\eeq
where $j=1,2,3$, $k=1,..,n_s$. In general, all of
these matrices are complex. Recall that the anticommutativity of fermion
fields in the path integral and the property $C^T=-C$ (where $C$ is the Dirac
charge conjugation matrix) imply that
\beq
M^{(f)} = M^{(f)T} \ \ , \ \  for \ \ f = L, R
\label{majsym}
\eeq
so that
\beq
M^{(N)} = M^{(N)T}
\label{m0sym}
\eeq
Thus, the situation in the leptonic sector is qualitatively
more complicated than that in the quark sector because of the general
presence of three types of neutral fermion bilinears, Dirac, left-handed
Majorana,
and right-handed Majorana, each with its own gauge and rephasing properties.
The diagonalization of $M^{(\ell)}$ yields the three charged lepton mass
eigenstates $e_m$, $\mu_m$, and $\tau_m$, while the diagonalization of
$M^{(N)}$ yields, in general, $3+n_s$ nondegenerate Majorana neutrino
mass eigenstates (some could, of course, be degenerate in magnitude, leading
to Dirac neutrino states).

    A comment is in order concerning possible electroweak-singlet neutrinos.
In the supersymmetric generalization of the standard model,
which is motivated by its
ability to maintain the condition $v << M_P$ beyond tree level without
fine-tuning, the component fields $\chi^c_{j,L}$ would arise as part of
$G_{SM}$-singlet chiral superfields $\hat \chi^c_{j}$ (all chiral superfields
will be written as left-handed).  One may recall that, in general, chiral
superfields which are singlets under the standard model gauge group $G_{SM}$
can destabilize the hierarchy \cite{args}.  However, in commonly used models,
the symmetry (such as matter parity) which prevents excessively rapid proton
decay also excludes the types of terms which would destabilize the hierarchy.

    To count the number of unremovable, and hence physically meaningful,
phases in $M^{(\ell)}$ and $M^{(N)}$, one rephases the lepton fields in
(\ref{ellmass}) and (\ref{numass}) so as to remove all possible phases in
these matrices.  We gave a general theorem on this counting problem in Ref.
\cite{ksl}.  To make our discussion reasonably self-contained, we briefly
review it here.  The rephasings are
\beq
L_{j L} = e^{ - i \alpha_j} L_{j L}'
\label{llrephase}
\eeq
\beq
\ell_{j R} = e^{i \beta^{(\ell)}_j} \ell_{j R}'
\label{lrrephase}
\eeq
for $j=1,2,3$, and, if there exist any $\chi_{j R}$'s,
also
\beq
\chi_{j R} = e^{i \beta^{(\chi)}_j} \chi_{j R}'
\label{chirephase}
\eeq
for $j=1,...n_s$.  In terms of the primed (rephased) lepton fields, the mass
matrices have elements
\beq
M_{jk}^{(\ell) \prime} = e^{i(\alpha_j + \beta^{(\ell)}_k)}M_{jk}^{(\ell)}
\label{mellrephased}
\eeq
for the charged leptons, and, for the neutrino sector,
\beq
M_{jk}^{(L) \prime} = e^{i(\alpha_j + \alpha_k)}M_{jk}^{(L)}
\label{mlrephased}
\eeq
for $j,k=1,2,3$;
\beq
M_{jk}^{(R) \prime} = e^{i(\beta^{(\chi)}_j + \beta^{(\chi)}_k)}M_{jk}^{(R)}
\label{mrrephased}
\eeq
for $j,k=1,...,n_s$, and
\beq
M_{jk}^{(D) \prime} = e^{i(\alpha_j + \beta^{(\chi)}_k)}M_{jk}^{(D)}
\label{mdrephased}
\eeq
for $j=1,2,3$ and $k=1,...,n_s$.

Thus, if $M^{(\ell)}$ has $N_\ell$ nonzero
(and, in general, complex) elements,
then the $N_\ell$ equations for making these elements real are
\beq
\alpha_j + \beta^{(\ell)}_k = -arg(M_{jk}^{(\ell)}) + \eta^{(\ell)}_{jk}\pi
\label{mellrephaseq}
\eeq
where the set $\{jk\}$ runs over each of these nonzero elements,
and $\eta^{(f)}_{jk} = 0$ or $1$.\footnote{The $\eta_{jk}$ term
allows for the possibility of making the
rephased element real and negative; this will not affect the counting of
unremovable phases.}

   Similarly, in the neutrino
sector, if $M^{(D)}$, $M^{(L)}$, and $M^{(R)}$ have, respectively,
$N_D$, $N_L$, and $N_R$ nonzero (and, in general, complex) elements, then the
corresponding equations for making these elements real are
\beq
\alpha_j + \beta^{(\chi)}_k = -arg(M_{jk}^{(D)}) + \eta^{(D)}_{jk}\pi
\label{mdrephaseq}
\eeq
\beq
\alpha_j + \alpha_k = -arg(M_{jk}^{(L)}) + \eta^{(L)}_{jk}\pi
\label{mlrephaseq}
\eeq
and
\beq
\beta^{(\chi)}_j + \beta^{(\chi)}_k = -arg(M_{jk}^{(R)}) + \eta^{(R)}_{jk}\pi
\label{mrrephaseq}
\eeq
where the various ranges of indices are obvious from (\ref{numass}).
Let us define the $(6+n_s)$-dimensional vector of fermion field phases
\beq
v = (\{\alpha_i \}, \{ \beta^{(\ell)}_i \}, \{\beta^{(\chi)}_i\})^T
\label{vvector}
\eeq
where $\{\alpha_i\} \equiv \{\alpha_1, \alpha_2, \alpha_3\}$,
$\{\beta^{(\ell)}_i\} \equiv
\{\beta^{(\ell)}_1, \beta^{(\ell)}_2, \beta^{(\ell)}_3\}$, and
$\{\beta^{(\chi)}_i\} \equiv
\{\beta^{(\chi)}_1,...,\beta^{(\chi)}_{n_s} \}$.
We also define the vector of phases of elements of the various mass
matrices\footnote{In eq. (\ref{wvector}) we change notation slightly from
Ref. \cite{ksl}, where an overall minus sign was absorbed in $T$.}
\beq
w=(\{-\arg(M^{(\ell)}_{jk})+\eta^{(\ell)}_{jk}\pi \},
\{\{-\arg(M^{(N)}_{mn})+\eta^{(N)}_{mn}\pi \})^T
\label{wvector}
\eeq
The dimension of the vector $w$ is equal to the number of rephasing
equations $N_{eq} = N_\ell+N_N$, where $N_N=N_L+N_R+N_D$.  The indices
$jk$ and $mn$ in (\ref{wvector}) run over the nonzero elements of $M^{(\ell)}$
and the independent nonzero elements of the symmetric matrix
$M^{(N)}$.  We can then write (\ref{mellrephaseq})-(\ref{mrrephaseq}) as
\beq
T v = w
\label{teq}
\eeq
which defines the matrix $T$. Since $T$ is an $N_{eq}$-row by
$(6+n_s)$-column matrix, clearly
\beq
rank(T) \le \min \{ N_{eq}, \ 6+n_s \}
\label{rt1}
\eeq
Recall that in the quark case \cite{ksq}, the analogous inequality was
$rank(T) \le \min\{N_{eq}, \ 9 \}$, but for realistic models where
$N_{eq} > 9$, one had the stronger inequality that $rank(T) \le 8$
because one overall rephasing left the Yukawa (or equivalently, mass)
matrices invariant.  In contrast, in the corresponding
present case, where $N_{eq} > 6+n_s$,
there is no overall rephasing (unless $M^{(L)}$ and
$M^{(R)}$ both vanish) which leaves $M^{(\ell)}$ and $M^{(N)}$ invariant, and
hence the inequality may be realized as an equality, i.e., $rank(T)$ may be
equal to $6+n_s$.  This will be illustrated in specific models to be
discussed in section 4.

   The first main theorem is then \cite{ksl}:
The number of unremovable and hence physically
meaningful phases $N_p$ in $M^{(\ell)}$ and $M^{(N)}$ is
\beq
N_p = N_{eq}- rank(T)
\label{np}
\eeq
For the proof, see Ref. \cite{ksl}. As noted there, this is independent of
whether or not the matrices are initially (complex) symmetric.  There is a
one-to-one correspondence between each such unremovable phase and the
phase of a rephasing-invariant
product of elements of mass matrices.  We discuss these invariants next.

\section{Rephasing Invariants and Theorems on Locations of Phases}
\label{invariants}

 A fundamental question concerns which elements of $M^{(\ell)}$ and $M^{(N)}$
can be made real by rephasings of lepton fields.  In Ref. \cite{ksl} several
theorems were given which provide a general answer to these questions.
The general method is to construct all independent complex products of
elements of the $M^{(f)}$, $f=\ell,N$, having the property that these products
are invariant under the rephasings (\ref{llrephase})-(\ref{chirephase}).
We refer to these as complex invariant products.  Since in general, by
construction, these invariant products are complex, each one implies a
constraint, which is that the set of elements which comprise it cannot all
be made real by the rephasings (\ref{llrephase})-(\ref{chirephase}) of the
lepton fields.
We define an irreducible complex invariant to be one which cannot be
factorized purely into products of lower-order complex invariants.  Since
reducible complex invariants do not yield any new phase constraints, it
suffices to consider only irreducible complex invariants, and we shall do so.
We define a set of
independent (irreducible) complex invariants to be a set of complex
(irreducible) invariants with the property that no invariant in the set is
equal to (i) the complex conjugate of another invariant in the set or (ii)
another element in the set with its indices permuted.  This does not imply that
the arguments of a set of independent (irreducible) complex invariants are
linearly independent.  We define $N_{ia}$ to
denote the number of linearly independent arguments among
the independent complex invariants.  Further, we define $N_{inv}$
to be the total number of independent (irreducible) complex invariants in a
given model.  It is useful to define $N_{inv,2n}$ to denote the number of
independent complex invariants of order $2n$.

    In Ref. \cite{ksl} we discussed the general construction of complex
rephasing invariants, enumerated the various types of these invariants at
fourth and sixth order, and gave explicit expressions for the fourth order
and a number of sixth order complex invariants.  Here we shall give a general
set of explicit expressions for complex invariants up to sixth order,
 inclusive. We recall our classification
of rephasing invariants into two general types:
(1) $P$-type, which involve elements of mass matrices in one charge sector
(charged leptons or neutrinos), and (2) $Q$-type, which involve elements of
mass matrices in both of these charge sectors.  We recall that the two
quartic $P$-type invariants are
\beq
P^{(f)}_{4;j_1 k_1, j_2 k_2; \tau} \equiv P^{(f)}_{j_1 k_1, j_2 k_2} =
M^{(f)}_{j_1 k_1}M^{(f)}_{j_2 k_2}M^{(f)*}_{j_2 k_1}M^{(f)*}_{j_1 k_2}
\label{p}
\eeq
for $f=\ell,N$. Note the symmetries
\beq
P^{(f)}_{j_1 k_1, j_2 k_2}=P^{(f)}_{j_2 k_2, j_1 k_1}
\label{pswitch}
\eeq
\beq
P^{(f)}_{j_1 k_1 j_2 k_2}=P^{(f)*}_{j_1 k_2, j_2 k_1}
\label{pconjugate}
\eeq
(whence also $P^{(f)}_{j_1 k_1, j_2 k_2} = P^{(f)*}_{j_2 k_1, j_1 k_2}$) for
$f=\ell, N$ and, for $f=N$, the additional symmetry
\beq
P^{(N)}_{j_1 j_2, j_3 j_4} = P^{(N)}_{j_2 j_1, j_4 j_3}
\label{p0sym}
\eeq
In a convenient shorthand notation, the complex invariants in (\ref{p}) for
$f=\ell,N$ may be denoted as $\ell \ell \ell^* \ell^*$ and $NNN^*N^*$,
respectively.
The $P_4$ complex invariants of the form $NNN^*N^*$ may be further
classified into six subtypes according to which submatrices in
$M^{(N)}$ they involve, $M^{(L)}$, $M^{(R)}$, and/or $M^{(D)}$:
\beqs
N N N^* N^*: & \quad L L L^* L^* & \ \qquad L D L^* D^* \nonumber \\
             & \quad R R R^* R^* & \ \qquad R D R^* D^* \nonumber \\
             & \quad D D D^* D^* & \ \qquad L R D^* D^*
\label{p4n}
\eeqs
(It is easily seen that the rephasing invariants of the subtype
$LRL^*R^*$ are real.)
Explicitly, the first three complex invariants are given by (\ref{p}) for
$f=L,R,D$, respectively, and the other three by the products
\beq
\Pi^{(L D)}_{j_1 j_2, j_3 k_1} =
M^{(L)}_{j_1 j_2}M^{(D)}_{j_3 k_1}M^{(L)*}_{j_1 j_3}M^{(D)*}_{j_2 k_1}
\label{rlf}
\eeq

\beq
\Xi^{(R D)}_{k_1 k_2, j_1 k_3} =
M^{(R)}_{k_1 k_2}M^{(D)}_{j_1 k_3}M^{(R)*}_{k_1 k_3}M^{(D)*}_{j_1 k_2}
\label{srd}
\eeq
and
\beq
\Omega^{(LRDD)}_{j_1 j_2, k_1 k_2} = M^{(L)}_{j_1 j_2}M^{(R)}_{k_1 k_2}
M^{(D)*}_{j_1 k_1}M^{(D)*}_{j_2 k_2}
\label{tlrdd}
\eeq
It is straightforward to express these in terms of the elements of
$P^{(N)}$, using (\ref{ml})-(\ref{md}):
\beq
P^{(L)}_{j_1 j_2, j_3 j_4} = P^{(N)}_{j_1 j_2, j_3 j_4}
\label{lllblbrel}
\eeq

\beq
P^{(R)}_{k_1 k_2, k_3 k_4} = P^{(N)}_{k_1+3 \ k_2+3, \ k_3+3 \ k_4+3}
\label{rrrbrbrel}
\eeq

\beq
P^{(D)}_{j_1 k_1, j_2 k_2} = P^{(N)}_{j_1 k_1+3, \ j_2 k_2+3}
\label{dddbdbrel}
\eeq

\beq
\Pi^{(L D)}_{j_1 j_2, j_3 k_1} = P^{(N)}_{j_2 \ j_1, \ j_3 \ k_1+3}
\label{pirel}
\eeq

\beq
\Xi^{(R D)}_{k_1 k_2, j_1 k_3} = P^{(N)}_{k_1+3 \ k_2+3, \ j_1 \ k_3+3}
\label{xirel}
\eeq

\beq
\Omega^{(LRDD)}_{j_1 j_2, k_1 k_2} =  P^{(N)}_{j_2 \ j_1, \ k_1+3 \ k_2+3}
\label{omrel}
\eeq

  At quartic order, we found two $Q$-type complex invariants, namely
\beq
Q^{(D \ell)}_{j_1 k_1,j_2 m_1} =
M^{(D)}_{j_1 k_1}M^{(\ell)}_{j_2 m_1}M^{(D)*}_{j_2 k_1}M^{(\ell)*}_{j_1 m_1}
\label{delldbellb}
\eeq
and
\beq
\Pi^{(L \ell)}_{j_1 j_2, j_3 m_1} =
M^{(L)}_{j_1 j_2}M^{(\ell)}_{j_3 m_1}
M^{(L)*}_{j_1 j_3}M^{(\ell)*}_{j_2 m_1}
\label{lelllbellb}
\eeq
In the short notation, these are denoted
$D \ell D^* \ell^*$ and $L \ell L^* \ell^*$. Note that
\beq
\Pi^{(L f)}_{j_1 j_2, j_3 m_1} = Q^{(L,f)}_{j_2 j_1, j_3 m_1}
\label{rq}
\eeq

    At sixth order there are first the complex invariants each of which
involves only one charge sector, $f f f f^* f^* f^*$ for $f=\ell,N$.  The
explicit expressions for these are
\beq
P^{(f)}_{j_1 k_1, j_2 k_2, j_3 k_3} =
M^{(f)}_{j_1 k_1}M^{(f)}_{j_2 k_2}M^{(f)}_{j_3 k_3}
M^{(f)*}_{j_2 k_1}M^{(f)*}_{j_3 k_2}M^{(f)*}_{j_1 k_3}
\label{p6}
\eeq
for $f=\ell,N$.  For any $f$, the invariants in (\ref{p6}) satisfy
\beq
P^{(f)}_{j_1 k_1, j_2 k_2, j_3 k_3} = P^{(f)}_{j_2 k_2, j_3 k_3, j_1 k_1} =
P^{(f)}_{j_3 k_3, j_1 k_1, j_2 k_2}
\label{p6cycl}
\eeq
and
\beq
P^{(f)}_{j_1 k_1,j_2 k_2, j_3 k_3} = P^{(f)*}_{j_3 k_2, j_2 k_1, j_1 k_3}
\label{p6sym}
\eeq
For the cases where $M^{(f)}$ is symmetric, namely $M^{(f)}=M^{(N)}$ or, for
submatrices, $M^{(L)}, M^{(R)}$, one has the additional symmetry
\beq
P^{(f)}_{j_1 j_2, j_3 j_4, j_5 j_6}=P^{(f)}_{j_2 j_1, j_6 j_5, j_4 j_3}
\label{p6maj}
\eeq
The $NNNN^*N^*N^*$ invariants may be divided into 11 subtypes according to
which submatrices of $M^{(N)}$ they involve:
\beqs
NNNN^*N^*N^*: & \quad LLLL^*L^*L^* & \quad\quad L L D L^* L^* D^* \nonumber \\
              & \quad RRRR^*R^*R^* & \quad\quad R R D R^* R^* D^* \nonumber \\
              & \quad DDDD^*D^*D^* & \quad\quad L R D L^* R^* D^* \nonumber \\
              & \quad DDLD^* D^* L^* & \quad\quad LLRL^* D^* D^* \nonumber \\
              & \quad DDRD^* D^* R^* & \quad\quad LRRD^* D^* R^* \nonumber \\
              & \quad & \quad\quad L R D D^* D^* D^*
\label{p60}
\eeqs
Complex conjugates of these invariants yield the same phase constraints and
hence are not listed.  Further, in the shorthand notation, the ordering of the
factors is not important; e.g. $DDLD^*D^*L^*$, $DLDD^*L^*D^*$, and
$LDDL^*D^*D^*$ represent the same subtype, and so forth for the others.
Note that the 6'th order rephasing invariants
$\ell R R \ell^* R^* R^*$,
$\ell \ell R \ell^* \ell^* R^*$,
$L R R L^* R^* R^*$,
$L L R L^* L^* R^*$, and
$\ell L R \ell^* L^* R^*$ all reduce to real factors times quartic complex
invariants and hence are not irreducible 6'th order complex invariants.  Given
an explicit formula for a particular type of $NNNN^*N^*N^*$ complex invariant
in terms of submatrices of $M^{(N)}$, it is straightforward to use eqs.
(\ref{ml})-(\ref{md}) to reexpress it in terms of $M^{(N)}$.
Thus, for $LLLL^*L^*L^*$, we have the relation
$P^{(L)}_{j_1 j_2, j_3 j_4, j_5 j_6} =
P^{(N)}_{j_1 j_2, j_3 j_4, j_5 j_6}$; for $DDDD^*D^*D^*$ the relation
$P^{(D)}_{j_1 k_1, j_2 k_2, j_3 k_3} =
P^{(N)}_{j_1 k_1+3, j_2 k_2+3, j_3 k_3+3}$, and so forth for the others.

Explicit expressions for the other subtypes of $NNNN^*N^*N^*$ complex
invariants in (\ref{p60}) are given below (in some cases there is more than
one kind of structure for a given subtype):
\beqs
DDLD^*D^*L^*: & \qquad
M^{(D)}_{j_1 k_1}M^{(D)}_{j_2 k_2}M^{(L)}_{j_3 j_4}
M^{(D)*}_{j_1 k_2}M^{(D)*}_{j_3 k_1}M^{(L)*}_{j_2 j_4}
\label{ddldbdblb1} \\
& \nonumber \\
& \qquad M^{(D)}_{j_1 k_1}M^{(D)}_{j_2 k_2}M^{(L)}_{j_3 j_4}
M^{(D)*}_{j_3 k_1}M^{(D)*}_{j_4 k_2}M^{(L)*}_{j_1 j_2}
\label{ddldbdblb2}
\eeqs

\beqs
DDRD^*D^*R^*: & \qquad
M^{(D)}_{j_1 k_1}M^{(D)}_{j_2 k_2}M^{(R)}_{k_3 k_4}
M^{(D)*}_{j_1 k_2}M^{(D)*}_{j_2 k_3}M^{(R)*}_{k_1 k_4}
\label{ddrdbdbrb1} \\
& \nonumber \\
& \qquad M^{(D)}_{j_1 k_1}M^{(D)}_{j_2 k_2}M^{(R)}_{k_3 k_4}
M^{(D)*}_{j_1 k_3}M^{(D)*}_{j_2 k_4}M^{(R)*}_{k_1 k_2}
\label{ddrdbdbrb2}
\eeqs

\beq
LLDL^*L^*D^*: \qquad
M^{(L)}_{j_1 j_2}M^{(L)}_{j_3 j_4}M^{(D)}_{j_5 k_1}
M^{(L)*}_{j_1 j_4}M^{(L)*}_{j_2 j_5}M^{(D)*}_{j_3 k_1}
\label{lldlblbdb}
\eeq

\beq
RRDR^*R^*D^*: \qquad
M^{(R)}_{k_1 k_2}M^{(R)}_{k_3 k_4}M^{(D)}_{j_1 k_5}
M^{(R)*}_{k_1 k_4}M^{(R)*}_{k_2 k_5}M^{(D)*}_{j_1 k_3}
\label{rrdrbrbdb}
\eeq

\beq
LRDL^*R^*D^*: \qquad
M^{(L)}_{j_1 j_2}M^{(R)}_{k_1 k_2}M^{(D)}_{j_3 k_3}
M^{(L)*}_{j_1 j_3}M^{(R)*}_{k_1 k_3}M^{(D)*}_{j_2 k_2}
\label{lrdlbrbdb}
\eeq

\beq
LLRL^*D^*D^*: \qquad
M^{(L)}_{j_1 j_2}M^{(L)}_{j_3 j_4}M^{(R)}_{k_1 k_2}
M^{(L)*}_{j_1 j_3}M^{(D)*}_{j_2 k_1}M^{(D)*}_{j_4 k_2}
\label{llrlbdbdb}
\eeq

\beq
LRRD^*D^*R^*: \qquad
M^{(L)}_{j_1 j_2}M^{(R)}_{k_1 k_2}M^{(R)}_{k_3 k_4}
M^{(D)*}_{j_1 k_2}M^{(D)*}_{j_2 k_4}M^{(R)*}_{k_1 k_3}
\label{lrrdbdbrb}
\eeq
and
\beqs
LRDD^*D^*D^*: & \qquad
M^{(L)}_{j_1 j_2}M^{(R)}_{k_1 k_2}M^{(D)}_{j_3 k_3}
M^{(D)*}_{j_1 k_3}M^{(D)*}_{j_2 k_2}M^{(D)*}_{j_3 k_1}
\label{lrddbdbdb1} \\
& \nonumber \\
& \qquad M^{(L)}_{j_1 j_2}M^{(R)}_{k_1 k_2}M^{(D)}_{j_3 k_3}
M^{(D)*}_{j_1 k_2}M^{(D)*}_{j_2 k_3}M^{(D)*}_{j_3 k_1}
\label{lrddbdbdb2}
\eeqs

Note that the invariants in (\ref{ddldbdblb1}) and
(\ref{lldlblbdb}) are of the form
\beq
Q^{(fff')}_{j_1 k_1, j_2 k_2, j_3 m_1} =
M^{(f)}_{j_1 k_1}M^{(f)}_{j_2 k_2}M^{(f')}_{j_3 m_1}
M^{(f)*}_{j_1 k_2}M^{(f)*}_{j_3 k_1}M^{(f')*}_{j_2 m_1}
\label{q6}
\eeq
for $(fff')=(DDL)$ and $(LLD)$, respectively.

We also find the following seven independent types of 6'th order complex
invariants linking the charge $Q=-1$ and $Q=0$ sectors:
\beqs
\ell \ell L \ell^* \ell^* L^* & \quad\quad \ell D L \ell^* D^* L^* \nonumber \\
L L \ell L^* L^* \ell^* & \quad\quad \ell D R \ell^* D^* R^* \nonumber \\
\ell \ell D \ell^* \ell^* D^* & \quad\quad \ell L R \ell^* D^* D^* \nonumber \\
D D \ell D^* D^* \ell^*
\label{qmqnlist}
\eeqs
The explicit forms for these are
\beqs
\ell \ell L \ell^* \ell^* L^*: \qquad
M^{(\ell)}_{j_1 m_1}M^{(\ell)}_{j_2 m_2}M^{(L)}_{j_3 j_4}
M^{(\ell)*}_{j_1 m_2}M^{(\ell)*}_{j_3 m_1}M^{(L)*}_{j_2 j_4}
\label{eelebeblb1} \\
& \qquad \nonumber \\
\qquad M^{(\ell)}_{j_1 m_1}M^{(\ell)}_{j_2 m_2}M^{(L)}_{j_3 j_4}
M^{(\ell)*}_{j_3 m_1}M^{(\ell)*}_{j_4 m_2}M^{(L)*}_{j_1 j_2}
\label{eelebeblb2}
\eeqs

\beq
LL\ell L^*L^*\ell^*: \qquad
M^{(L)}_{j_1 j_2}M^{(L)}_{j_3 j_4}M^{(\ell)}_{j_5 m_1}
M^{(L)*}_{j_1 j_4}M^{(L)*}_{j_5 j_2}M^{(\ell)*}_{j_3 m_1}
\label{llelblbeb}
\eeq

\beq
\ell\ell D \ell^*\ell^*D^*: \qquad
M^{(\ell)}_{j_1 m_1}M^{(\ell)}_{j_2 m_2}M^{(D)}_{j_3 k_1}
M^{(\ell)*}_{j_1 m_2}M^{(\ell)*}_{j_3 m_1}M^{(D)*}_{j_2 k_1}
\label{eedebebdb}
\eeq

\beq
\ell D R \ell^* D^* R^*: \qquad
M^{(\ell)}_{j_1 m_1}M^{(D)}_{j_2 k_1}M^{(R)}_{k_2 k_3}
M^{(\ell)*}_{j_2 m_1}M^{(D)*}_{j_1 k_2}M^{(R)*}_{k_1 k_3}
\label{edrebdbrb}
\eeq

\beq
\ell L R \ell^* D^* D^*: \qquad
M^{(\ell)}_{j_1 m_1}M^{(L)}_{j_2 j_3}M^{(R)}_{k_1 k_2}
M^{(\ell)*}_{j_2 m_1}M^{(D)*}_{j_1 k_1}M^{(D)*}_{j_3 k_2}
\label{elrebdbdb}
\eeq

\beqs
\ell D L \ell^* D^* L^*: & \qquad
M^{(\ell)}_{j_1 m_1}M^{(D)}_{j_2 k_1}M^{(L)}_{j_3 j_4}
M^{(\ell)*}_{j_2 m_1}M^{(D)*}_{j_4 k_1}M^{(L)*}_{j_3 j_1}
\label{edlebeblb_1} \\
& \nonumber \\ &
M^{(\ell)}_{j_1 m_1}M^{(D)}_{j_2 k_1}M^{(L)}_{j_3 j_4}
M^{(\ell)*}_{j_4 m_1}M^{(D)*}_{j_1 k_1}M^{(L)*}_{j_3 j_2}
\label{edlebeblb_2} \\
& \nonumber \\ &
M^{(\ell)}_{j_1 m_1}M^{(D)}_{j_2 k_1}M^{(L)}_{j_3 j_4}
M^{(\ell)*}_{j_2 m_1}M^{(D)*}_{j_3 k_1}M^{(L)*}_{j_1 j_4}
\label{edlebeblb_3} \\
& \nonumber \\ &
M^{(\ell)}_{j_1 m_1}M^{(D)}_{j_2 k_1}M^{(L)}_{j_3 j_4}
M^{(\ell)*}_{j_2 m_1}M^{(D)*}_{j_3 k_1}M^{(L)*}_{j_1 j_4}
\label{edlebeblb_4} \\
& \nonumber \\ &
M^{(\ell)}_{j_1 m_1}M^{(D)}_{j_2 k_1}M^{(L)}_{j_3 j_4}
M^{(\ell)*}_{j_3 m_1}M^{(D)*}_{j_4 k_1}M^{(L)*}_{j_1 j_2}
\label{edlebeblb_5} \\
& \nonumber \\ &
M^{(\ell)}_{j_1 m_1}M^{(D)}_{j_2 k_1}M^{(L)}_{j_3 j_4}
M^{(\ell)*}_{j_4 m_1}M^{(D)*}_{j_3 k_1}M^{(L)*}_{j_1 j_2}
\label{edlebeblb_6}
\eeqs
The methods for determining these are similar
to those discussed in detail for the quark case in Ref. \cite{ksl}.  For
example, the general explicit form of the $\ell D L \ell^* D^* L^*$ invariants
is
\beq
M^{(\ell)}_{j_1 m_1}M^{(D)}_{j_2 k_1}M^{(L)}_{j_3 j_4}
M^{(\ell)*}_{\sigma(j_1) m_1}M^{(D)*}_{\sigma(j_2) k_1}
M^{(L)*}_{\sigma(j_3) \sigma(j_4)}
\label{edlebeblbgen}
\eeq
where $\sigma \in S_4$, the group of permutations of 4 indices.  We then
determine the full subset of these $4!$ products which yields independent
irreducible sixth order complex invariants, and obtain the results listed in
(\ref{edlebeblb_1})-(\ref{edlebeblb_6}).  Note that (\ref{eelebeblb1}) and
(\ref{llelblbeb}) are again of the form (\ref{q6}) for $(fff')=(\ell\ell L)$
and $(LL\ell)$, respectively.

   In the case of quark mass matrices (with $N_G=3$ generations of quarks), we
showed\cite{ksq} that the fourth and sixth order complex invariants sufficed to
describe all of the phase constraints on the mass matrices.  An analogous
result does not hold in the leptonic sector, however; in general, it may be
necessary to include complex invariants of higher order than sixth to obtain
all of the phase constraints.  The reason for this is just that the neutral
part of the lepton sector is, in general, more complicated than either the up
or down quark sector.  Rather than enumerate in general all of the possible
eighth-order complex invariants, which would be quite involved, we will make
clear how one uses them in practical calculations via two specific models in
section 4.

    We next review our theorems on invariants, as applied to lepton mass
matrices.\cite{ksl}  For a given model, one constructs the maximal set of
$N_{ia}$ independent complex
invariants of lowest order(s), whose arguments (phases) are linearly
independent. Then (a) each of these invariants implies a constraint that the
elements contained within it cannot, in general, all be made
simultaneously real; (b) this constitutes the complete set of
constraints on which elements of $M^{(\ell)}$ and $M^{(N)}$ can be made
simultaneously real; and hence (c) $N_p = N_{ia}$.

   An immediate corollary is:
If in a given model there are as at least as many independent quartic
complex invariants with independent arguments as there are unremovable phases,
then the arguments of all higher order
complex invariants are expressible in terms of those of the quartic invariants
and hence yield no new phase constraints.

    In the quark case, we presented a useful graphical representation for the
complex invariants in Ref. \cite{ksq}.  This again works directly for
invariants involving only the charged lepton sector.  For invariants involving
the neutral lepton sector, there is no $1-1$ correspondence between a matrix
element $M^{(N)}_{jk}$ occuring in a given complex invariant and a point on
an graphical array representing the matrix $M^{(N)}$; instead, as a result of
the symmetry (\ref{m0sym}), $M^{(N)T} = M^{(N)}$, there is a obvious 2-fold
homomorphism according to which the element $M^{(N)}_{jk}=M^{(N)}_{kj}$
corresponds to the
two points ($j$'th row, $k$'th column) and ($k$'th row, $j$'th column) in the
array representing $M^{(N)}$.  For an invariant involving $2n$ elements of
$M^{(N)}$ of which a (possibly null) subset of $r$ are diagonal elements, there
are $2^{2n-k}$-ways of representing it graphically.  Among these ways, however,
one can always find a graphical representation which is analogous to those for
the quark sector complex invariants discussed in Ref. \cite{ksq}.
Moreover, as a tool for quickly determining the complex invariants, given the
forms of the mass matrices, the graphical method is equally useful in the
leptonic case as in the quark case.  We shall illustrate it in this context
below.

   Since the full set of $N_{inv}$ independent complex invariants
will have arguments which are not, in general, linearly independent, it follows
that
\beq
N_{inv} \ge N_{ia}
\label{ninvia}
\eeq
It may also happen that, e.g. for order $2n=4$, the number of independent
quartic complex invariants, $N_{inv,4}$ is greater than $N_{ia}$.  Using the
methods of Ref. \cite{ksl} (see also Ref. \cite{ksq}), one can determine how
many of the complex invariants of all orders, or of a particular order, have
linearly independent arguments, and can thus construct sets of such invariants
with this property.

\section{Applications to Specific Models}
\label{models}

    Although our theorems are quite general, it is useful to see how they apply
to various specific models.  In Ref. \cite{ksl} we gave a brief discussion of
applications.  Here we will consider several more models.
In contrast to the quark sector, where one has reasonably
accurate measurements of most Cabibbo-Kobayashi-Maskawa (CKM) quark
mixing matrix elements, in the leptonic sector, there is not even any definite
direct evidence of leptonic mixing, let alone accurate measurements of
various mixing matrix elements.  As noted above, probably the strongest
indirect evidence comes from the apparent deficiency of the solar neutrino
flux.  The situation concerning possible evidence for atmospheric neutrino
oscillations is unclear at present \cite{pdg,nurev}.  We shall tentatively
assume that the solar neutrino deficit does indicate neutrino
oscillations.\footnote{For an recent phenomenological study which also includes
fits to atmospheric neutrino oscillations, see, e.g., Ref. \cite{albright}.}
Moreover, since our purpose is to illustrate the
application of our general results on phases, we will not discuss the details
of the phenomenology of the models.

   In general, models for fermion masses and mixing
can be classified according to whether they assume a theoretical
framework of perturbative or nonperturbative electroweak symmetry
breaking.  We shall concentrate here on the class of models in which
the observed electroweak symmetry breaking is perturbative.  In this class,
an appealing framework is provided by supersymmetric extensions of the
standard model, which stabilize the Higgs sector and hence stabilize the
hierarchy between the electroweak scale and the Planck scale.\cite{susy}
In these models,
one typically hypothesizes some simple forms at a high mass scale (e.g., a mass
scale characterizing a possible grand unified theory (GUT) or a mass scale
near to the string scale, $\bar M_P$) for the Yukawa and higher-dimensional
operators which are responsible for fermion mass generation.  As part of
this, one assumes some symmetries to prevent various Yukawa couplings and
higher-dimensional operators, and thereby render various entries in the
(effective) Yukawa matrices zero.  The purpose of this is, of course, to
minimize the number of parameters and hence increase the apparent
predictiveness.  One then evolves these forms for the Yukawa matrices down to
the electroweak mass scale using the appropriate renormalization group
equations.  For this evolution, we use the renormalization group equations
of the minimal supersymmetric standard model (MSSM).  Parenthetically,
we recall that when one evolves these forms for the Yukawa
matrices down to the electroweak level, the zero entries do not, in general,
remain zero.  Clearly, the entries which are modelled as being exactly zero
might well be nonzero, but small quantities.  For example, one possibility is
that they are suppressed by certain powers of $(M_r/\bar M_P)$, where $M_r$ is
the reference scale at which one analyzes the effective Yukawa interaction.
Since our results do not depend on whether or
not the mass matrices are symmetric at some mass scale, we shall consider the
general case of non-symmetric mass matrices.

\subsection{Model Without Electroweak-Singlet Neutrinos}

    We first consider a model with no electroweak-singlet neutrinos, i.e.,
with $n_s=0$.  This is defined by \cite{ksl}.
\beq
M^{(\ell)} =  \left (\begin{array}{ccc}
                  0 & E_{12} & 0 \\
                  E_{21} & E_{22} & E_{23} \\
                  0 &  E_{32} &  E_{33} \end{array}   \right  )
\label{me1}
\eeq

\beq
M^{(L)} =  \left (\begin{array}{ccc}
                  0 & L_{12} & 0 \\
                  L_{12} & L_{22} & L_{23} \\
                  0 &  L_{23} &  L_{33} \end{array}   \right  )
\label{ml1}
\eeq
(where each element is, in general, complex, since no symmetry forces it to be
real).  We have verified that, for appropriate choices of the parameters,
 this model is able to fit current solar
neutrino data and other established limits on neutrino masses and mixing.
For this model, there are $N_{eq}=10$ complex elements in $M^{(\ell)}$ and
$M^{(D)}$ and corresponding rephasing equations.  We calculate the matrix $T$
and find $rank(T)=6$.  Our theorem (\ref{np}) then implies that
there are $N_p=N_{eq}-rank(T)=4$ unremovable phases in $M^{(\ell)}$ and
$M^{(L)}$.   (This model provides an illustration of how
$rank(T)$ may realize the inequality (\ref{rt1}) as an equality.)
We find $N_{inv,4}=8$ independent complex (quartic) invariants:
\beq
P^{(\ell)}_{22,33}=E_{22}E_{33}E_{32}^*E_{23}^*
\label{pe2233}
\eeq

\beq
P^{(L)}_{22,33}=L_{22}L_{33}L_{23}^{*2}
\label{pl2233}
\eeq

\beq
\Pi^{(L \ell)}_{12,22}=L_{12}E_{22}L_{22}^*E_{12}^*
\label{lele1222}
\eeq

\beq
\Pi^{(L \ell)}_{12,32}=L_{12}E_{32}L_{23}^*E_{12}^*
\label{lele12322}
\eeq

\beq
\Pi^{(L \ell)}_{22,32}=L_{22}E_{32}L_{23}^*E_{22}^*
\label{lele2232}
\eeq

\beq
\Pi^{(L \ell)}_{22,33}=L_{22}E_{33}L_{23}^*E_{23}^*
\label{lele2233}
\eeq

\beq
\Pi^{(L \ell)}_{23,32}=L_{23}E_{32}L_{33}^*E_{22}^*
\label{lele2332}
\eeq

\beq
\Pi^{(L \ell)}_{23,33}=L_{23}E_{33}L_{33}^*E_{23}^*
\label{lele2333}
\eeq
 From these we calculate the corresponding $Z$ matrix and find
that it has rank 4, so that of the eight arguments of the complex invariants,
there are $N_{ia}=4$ linearly independent ones, in accord with the equality
$N_{ia}=N_p$ and our result that $N_p=4$. There are correspondingly four
independent phase constraints.  In particular, for this model it is not,
in general, possible by any rephasings to make either the charged lepton or
neutrino mass matrices real.  A set of complex invariants with linearly
independent arguments is given by
\beq
\{P^{(\ell)}_{22,33}, \ P^{(L)}_{22,33}, \ \Pi^{(L \ell)}_{12,22}, \
\Pi^{(L \ell)}_{22,32} \}
\label{quarticset1}
\eeq
We denote the arguments of the invariants in eq. (\ref{quarticset1}
as $\theta_j$, $j=1,...,4$; i.e.,
$\theta_1=\arg(P^{(\ell)}_{22,33})$, etc. The only element which does not
appear in any complex invariant and hence may be rephased freely is $E_{21}$.

   A necessary task which a model builder must carry out is to determine which
of the elements of the mass matrices can be rendered real by the rephasing
of the lepton fields.  We display below an example of an allowed form for
the mass matrices after such rephasing is
\beq
M^{(\ell)'} =  \left (\begin{array}{ccc}
       0 & |E_{12}|e^{-i\theta_3} & 0 \\
      |E_{21}| & |E_{22}| & |E_{23}| \\
      0 &  |E_{32}|e^{i\theta_4} &
      |E_{33}|e^{i(\theta_1+\theta_4)} \end{array} \right )
\label{me2'}
\eeq

\beq
M^{(L)'} =  \left (\begin{array}{ccc}
               0 & |L_{12}| & 0 \\
            |L_{12}| & |L_{22}| & |L_{23}| \\
 0 &  |L_{23}| &  |L_{33}|e^{i\theta_2} \end{array} \right )
\label{ml2'}
\eeq

This model has $N_{inv,6}=4$ sixth order complex invariants:
\beq
Q^{(LL\ell)}_{22,33,12} = L_{22}L_{33}E_{12}L_{23}^*L_{12}^*E_{32}^*
\label{qlle223312}
\eeq

\beq
Q^{(\ell\ell L)}_{22,33,12} = E_{22}E_{33}L_{12}E_{23}^*E_{12}^*L_{23}^*
\label{qeel223312}
\eeq

\beq
Q^{(LL\ell)}_{33,12,22} = L_{33}L_{12}E_{22}L_{23}^{*2}E_{12}^*
\label{qlle331222}
\eeq

\beq
Q^{(\ell\ell L)}_{33,12,22} = E_{33}E_{12}L_{22}E_{32}^*E_{23}^*L_{12}^*
\label{qeel331222}
\eeq
However, as a result of our corollary 1 above, since we have already exhibited
a set of quartic complex invariants with independent arguments and comprised of
$N_{ia}$ members equal to the number $N_p$ of unremovable phases in the mass
matrices, it follows that all higher order invariants, and, in particular, the
sixth order ones listed above, have arguments which can be expressed as linear
combinations of those of the set of $N_{ia}$ fourth order invariants and hence
do not yield any new phase constraints.

\subsection{Models with Electroweak-Singlet Neutrinos}

\subsubsection{$n_s=3$ Model 1}

    We will concentrate on models with $n_s=N_G=3$.  Our methods are easily
applied to models with other values of $n_s$.  We begin with an idealized model
in which left-handed Majorana mass terms are assumed to be negligible, i.e.,
$M^{(L)}=0$, and the other mass matrices are given by
\beq
M^{(\ell)} =  \left (\begin{array}{ccc}
                  0 & E_{12} & 0 \\
                  E_{21} & E_{22} & E_{23} \\
                  0 &  E_{32} &  E_{33} \end{array}   \right  )
\label{mes1}
\eeq

\beq
M^{(R)} =  \left (\begin{array}{ccc}
                  R_{11} & 0 & 0 \\
                  0 & R_{22} & R_{23} \\
                  0 &  R_{23} &  R_{33} \end{array}   \right  )
\label{mrs1}
\eeq

\beq
M^{(D)} =  \left (\begin{array}{ccc}
                  0 & D_{12} & 0 \\
                  D_{21} & 0 & D_{23} \\
                  0 &  D_{32} &  D_{33} \end{array}   \right  )
\label{mds1}
\eeq
We have again checked that with appropriate choices of the parameters, this
model can fit the experimental constraints discussed at the beginning of the
paper. The model has $N_{eq}=15$, and we compute $rank(T)=9$, so that there
are $N_p=6$ unremovable phases in the mass matrices.  The model has
$N_{inv,4}=6$ quartic complex invariants:
\beq
P^{(\ell)}_{22,33}=E_{22}E_{33}E_{23}^*E_{32}^*
\label{pe2233a}
\eeq

\beq
Q^{(D\ell)}_{23,33}=D_{23}E_{33}D_{33}^*E_{23}^*
\label{dedbeb2333}
\eeq

\beq
Q^{(D\ell)}_{23,32}=D_{23}E_{32}D_{33}^*E_{22}^*
\label{dedbeb2332}
\eeq

\beq
Q^{(D\ell)}_{12,32}=D_{12}E_{32}D_{32}^*E_{12}^*
\label{dedbeb1232}
\eeq

\beq
\Xi^{(RD)}_{23,33}=R_{23}D_{33}R_{33}^*D_{32}^*
\label{rdrbdb2333}
\eeq
and
\beq
P^{(R)}_{22,33}=R_{22}R_{33}R_{23}^{*2}
\label{pr2233}
\eeq
Using our $Z$ matrix method, we find that among these six invariants there
are five linearly independent arguments.  A set of five quartic complex
invariants with linearly independent arguments is provided by
two among the subset $\{P^{(\ell)}_{22,33}, \ Q^{(D\ell)}_{23,33}, \
Q^{(D\ell)}_{23,32}\}$ together with the three invariants
$Q^{(D\ell)}_{12,32}$, $\Xi^{(RD)}_{23,33}$, and $P^{(R)}_{22,33}$.  To obtain
an invariant encoding the sixth unremovable phase, we consider 6'th order
complex products.  We find $N_{inv,6}=7$ of these:
\beq
I_{6a}=E_{12}E_{23}D_{32}E_{22}^*E_{33}^*D_{12}^*
\label{eed122332}
\eeq

\beq
I_{6b}=D_{23}D_{32}E_{12}D_{12}^*D_{33}^*E_{22}^*
\label{dde233212}
\eeq

\beq
I_{6c}=E_{12}D_{23}R_{22}E_{22}^*D_{12}^*R_{23}^*
\label{edr233212}
\eeq

\beq
I_{6d}=E_{12}D_{33}R_{22}E_{32}^*D_{12}^*R_{23}^*
\label{edr123322}
\eeq

\beq
I_{6e}=E_{12}D_{23}R_{23}E_{22}^*D_{12}^*R_{33}^*
\label{edr122323}
\eeq

\beq
I_{6f}=E_{12}D_{33}R_{23}E_{32}^*D_{12}^*R_{33}^*
\label{edr123323}
\eeq
and
\beq
I_{6g}=D_{23}^2R_{11}D_{21}^{*2}R_{33}^*
\label{ddr232311}
\eeq
Among these seven 6'th order invariants, the arguments of the first six are
linearly
dependent upon the five arguments from the quartic invariants, but the
argument of $I_{6g}$ is linearly independent of these and hence accounts for
the sixth unremovable phase in the model. Therefore
a complete set of complex invariants is given by
\beq
\{ Q^{(D\ell)}_{23,33}, \ Q^{(D\ell)}_{23,32}, \ Q^{(D\ell)}_{12,32}, \
\Xi^{(RD)}_{23,33}, \ P^{(R)}_{22,33}, \ I_{6g} \}
\label{setns1}
\eeq
We denote the arguments of these complex invariants as $\phi_j$, $j=1,...,6$,
i.e., $\phi_1 = \arg(Q^{(D\ell)}_{23,32})$, etc.  An example of an allowed
form for the mass matrices in this model after rephasing of the lepton fields
is
\beq
M^{(\ell)'} =  \left (\begin{array}{ccc}
              0 & |E_{12}|e^{i(\phi_2-\phi_3)} & 0 \\
              |E_{21}| & |E_{22}| & |E_{23}| \\
     0 & |E_{32}|e^{i\phi_2} & |E_{33}|e^{i\phi_1} \end{array}  \right )
\label{mens1'}
\eeq

\beq
M^{(R)'} =  \left (\begin{array}{ccc}
        |R_{11}|e^{i\phi_6} & 0 & 0 \\
        0 & |R_{22}|e^{i(\phi_5+2\phi_4)} & |R_{23}|e^{i\phi_4} \\
        0 & |R_{23}|e^{i\phi_4} &  |R_{33}| \end{array}   \right  )
\label{mrns1'}
\eeq
and
\beq
M^{(D)'} =  \left (\begin{array}{ccc}
                  0 & |D_{12}| & 0 \\
                  |D_{21}| & 0 & |D_{23}| \\
                  0 &  |D_{32}| &  |D_{33}| \end{array}   \right  )
\label{mds1'}
\eeq

By considering a special case of this model, we can illustrate our statement in
section 3 that for leptonic mass matrices, in contrast to quark mass matrices,
the quartic and sixth order complex invariants do not necessarily suffice to
encode all of the unremovable phases.  For this purpose, let us, for example,
set $R_{33}=0$.  In this case,  $N_{eq}=14$ and the rank of the resultant
$T$ is again 9, so that $N_p=5$.  Among the quartic invariants, the first
four remain, but the last two, $\Xi^{(RD)}_{23,33}$ and $P^{(R)}_{22,33}$,
evidently vanish.  From our previous analysis it follows that
among the remaining four quartic invariants, there are three linearly
independent arguments.  Of the original seven sixth-order complex invariants,
the last three evidently vanish.  Since the arguments of the first four sixth
order invariants are all linearly dependent upon those of the quartic
invariants, they does not account for either of the two
remaining unremovable phases.  Consequently, one must proceed to construct the
complex 8'th order complex invariants for this special case.  We find the
following two:
\beq
I_{8a}=E_{22}D_{23}D_{32}R_{11}E_{32}^*D_{21}^{*2}R_{23}^*
\label{i8a}
\eeq

\beq
I_{8b}=E_{23}D_{23}D_{32}R_{11}E_{33}^*D_{21}^{*2}R_{23}^*
\label{i8b}
\eeq
The arguments of these two invariants are linearly independent of the three
arguments from the quartic invariants, yielding the full set of $N_{ia}=N_p=5$
independent arguments for this special case of the model where $R_{33}=0$.
A set of complex invariants with linearly independent phases is provided by
(i) three out of eqs. (\ref{pe2233a})-(\ref{pr2233}) together with (ii)
$I_{8a}$ and $I_{8b}$.  (The general model with $R_{33} \ne 0$ has
several additional 8'th order complex invariants, but they do not yield any new
phase constraints.)

\subsubsection{$n_s=3$ Model 2}

A second model with $n_s=3$ electroweak-singlet neutrinos is defined by the
mass matrices
\beq
M^{(\ell)} =  \left (\begin{array}{ccc}
                  0 & E_{12} & 0 \\
                  E_{21} & 0 & E_{23} \\
                  0 &  E_{32} &  E_{33} \end{array}   \right  )
\label{mes2}
\eeq

\beq
M^{(L)} =  \left (\begin{array}{ccc}
                  0 & L_{12} & 0 \\
                  L_{12} & 0 & 0 \\
                  0 &  0 &  L_{33} \end{array}   \right  )
\label{mls2}
\eeq

\beq
M^{(R)} =  \left (\begin{array}{ccc}
                  0 & R_{12} & 0 \\
                  R_{12} & 0 & R_{23} \\
                  0 &  R_{23} &  R_{33} \end{array}   \right  )
\label{mrs2}
\eeq

\beq
M^{(D)} =  \left (\begin{array}{ccc}
                  0 & D_{12} & 0 \\
                  D_{21} & 0 & D_{23} \\
                  0 &  D_{32} &  D_{33} \end{array}   \right  )
\label{mds2}
\eeq

This model has $N_{eq}=15$.  We calculate the $ 15 \times 9$ matrix $T$ and
find $rank(T)=9$, so that, according to our theorem, there are $N_p=6$
unremovable phases in the leptonic mass matrices.  To determine which elements
can be made real, we construct a set of complex invariants which yields the
full set of phase constraints.  We find that there are $N_{inv,4}=8$ quartic
complex invariants:
\beq
Q^{(D\ell)}_{12,32} = D_{12}E_{32}D_{32}^*E_{12}^*
\label{qde1232}
\eeq

\beq
Q^{(D\ell)}_{23,33} = D_{23}E_{33}D_{33}^*E_{23}^*
\label{qde2333}
\eeq

\beq
\Xi^{(RD)}_{21,23} = R_{12}D_{23}R_{23}^*D_{21}^*
\label{rd1223}
\eeq

\beq
\Xi^{(RD)}_{33,32} = R_{33}D_{32}R_{23}^*D_{33}^*
\label{rd3332}
\eeq

\beq
\Xi^{(RD)}_{21,32} = R_{12}D_{32}R_{23}^*D_{21}^*
\label{rd1232}
\eeq

\beq
\Omega^{(LRDD)}_{12,12} = L_{12}R_{12}D_{21}^*D_{12}^*
\label{lrdd1212}
\eeq

\beq
\Omega^{(LRDD)}_{12,23} = L_{12}R_{23}D_{23}^*D_{12}^*
\label{lrdd1223}
\eeq

\beq
\Omega^{(LRDD)}_{33,23} = L_{33}R_{23}D_{32}^*D_{33}^*
\label{lrdd3323}
\eeq
By our usual $Z$ matrix method, we determine that among these eight invariants
there are six linearly independent arguments, so that the quartic complex
invariants suffice to yield all phase constraints in this model.  We find
that a subset of the eight quartic invariants which have independent
arguments is given, e.g., by
\beq
\{Q^{(D\ell)}_{12,32}, \ \ Q^{(D\ell)}_{23,33}, \ \ \Xi^{(RD)}_{33,32}, \ \
\Omega^{(LRDD)}_{12,12}, \ \ \Omega^{(LRDD)}_{12,23}, \ \
\Omega^{(LRDD)}_{33,23} \}
\label{quarticset}
\eeq
We denote the arguments of these complex invariants as $\omega_j$, $j=1,..,6$,
respectively. An example of an allowed set of mass matrices after rephasing is
\beq
M^{(\ell)} =  \left (\begin{array}{ccc}
      0 & |E_{12}| & 0 \\
      |E_{21}| & 0 & |E_{23}| \\
  0 &  |E_{32}|e^{i\omega_1} &  |E_{33}|e^{i\omega_2} \end{array}   \right  )
\label{mes2'}
\eeq

\beq
M^{(L)} =  \left (\begin{array}{ccc}
                  0 & |L_{12}| & 0 \\
                  |L_{12}| & 0 & 0 \\
                  0 &  0 &  |L_{33}| \end{array}   \right  )
\label{mls2'}
\eeq

\beq
M^{(R)} =  \left (\begin{array}{ccc}
     0 & |R_{12}|e^{i\omega_3} & 0 \\
     |R_{12}|e^{i\omega_3} & 0 & |R_{23}| \\
     0 &  |R_{23}| &  |R_{33}|e^{i\omega_4} \end{array}   \right  )
\label{mrs2'}
\eeq

\beq
M^{(D)} =  \left (\begin{array}{ccc}
     0 & |D_{12}| & 0 \\
     |D_{21}|e^{-i\omega_5} & 0 & |D_{23}|e^{-i\omega_6} \\
     0 &  |D_{32}| &  |D_{33}| \end{array}   \right  )
\label{mds2'}
\eeq
By our corollary 1, since we have constructed a set (\ref{quarticset})
of quartic invariants with
the full $N_{ia}=N_p$ set of linearly independent arguments, all higher order
complex invariants have arguments which can be expressed in terms of those of
this set, and hence they do not yield any new phase constraints.
For completeness, we exhibit the 6'th order complex invariants in this model.
Two of these are of $LRDD^*D^*D^*$ type,
\beq
P^{(N)}_{12,66,35}=L_{12}R_{33}D_{32}D_{12}^*D_{23}^*D_{33}^*
\label{lrd123332}
\eeq

\beq
P^{(N)}_{26,33,54}=L_{33}R_{12}D_{23}D_{21}^*D_{32}^*D_{33}^*
\label{lrd331223}
\eeq
while the third is of $DDLD^*D^*L^*$ type:
\beq
P^{(N)}_{12,63,35}=D_{32}D_{33}L_{12}D_{12}^*D_{23}^*L_{23}^*
\label{ddl323312}
\eeq
(Several choices of indices on the $P^{(N)}_{j_1 k_1 j_2 k_2 j_3 k_3}$ yield
the same invariants; we have listed only one for each case.)

\subsubsection{$n_s=3$ Model 3}

   We next give two toy models to illustrate some theoretical points. The
first is
an example of a model with no complex quartic invariants, where the first
complex invariant occurs at sixth order.  It is defined by
\beq
M^{(\ell)} =  \left (\begin{array}{ccc}
                  E_{11} & 0 & 0 \\
                  0 & E_{22} & 0 \\
                  0 & 0 &  E_{33} \end{array}   \right  )
\label{mes3}
\eeq

\beq
M^{(L)} =  \left (\begin{array}{ccc}
                  L_{11} & 0 & 0 \\
                  0 & 0 & 0 \\
                  0 &  0 &  0 \end{array}   \right  )
\label{mls3}
\eeq

\beq
M^{(R)} =  \left (\begin{array}{ccc}
                  R_{11} & 0 & 0 \\
                  0 & R_{22} & 0 \\
                  0 &  0 &  R_{33} \end{array}   \right  )
\label{mrs3}
\eeq

\beq
M^{(D)} =  \left (\begin{array}{ccc}
                  0 & 0 & 0 \\
                  0 & 0 & D_{23} \\
                  0 &  D_{32} &  D_{33} \end{array}   \right  )
\label{mds3}
\eeq
We calculate that $rank(T)=9$ for this model, so that
there is $N_p=1$ unremovable phase.  The single independent complex sixth order
invariant is of $DDRD^*D^*R^*$ type:
\beq
D_{32}^2R_{33}D_{33}^{*2}R_{22}^*
\label{invmod3}
\eeq
The corresponding phase may be placed in any of the mass matrix elements
involved in (\ref{invmod3}).

\subsubsection{$n_s=3$ Model 4}

   This model illustrates that, in contrast to the case in the quark sector,
the quartic and sixth order complex invariants do not necessarily suffice to
yield all phase constraints in the leptonic sector.  The model is defined by
$M^{(\ell)}$ and $M^{(D)}$ as in eqs. (\ref{mes3}) and (\ref{mds3}),
respectively, together with
\beq
M^{(L)} =  \left (\begin{array}{ccc}
                  L_{11} & 0 & 0 \\
                  0 &  L_{22} & 0 \\
                  0 &  0 &  0 \end{array}   \right  )
\label{mls4}
\eeq
and
\beq
M^{(R)} =  \left (\begin{array}{ccc}
                  R_{11} & 0 & 0 \\
                  0 & R_{22} & 0 \\
                  0 &  0 &  0 \end{array}   \right  )
\label{mrs4}
\eeq
The $T$ matrix for this model has rank 9, so that $N_p=1$.
There are no (nonzero) complex quartic or sixth order invariants.  The model
has the single independent 8'th order invariant
\beq
L_{22}R_{22}D_{33}^2D_{23}^{*2}D_{32}^{*2}
\label{invmod4}
\eeq
As before, the corresponding phase may be placed in any of the mass matrix
elements involved in (\ref{invmod3}).

\subsection{Unified Models of Quark and Lepton Mass Matrices}

   One of the appeals of grand unified theories has always been the fact that
they allow one naturally to relate quark and lepton Yukawa, and hence mass,
matrices.\footnote{This appeal is offset somewhat by the fact that even
when one
considers supersymmetric grand unified theories to stablize the overall gauge
hierarchy, there is a still a serious problem associated with the necessity of
splitting the electroweak doublet Higgs fields from color triplet Higgs fields
which occur in the same representations of the GUT gauge group.  Furthermore,
the most likely source for the additional symmetries which model-builders
impose to get zeroes in the (effective) Yukawa matrices is an underlying string
theory.  But, at least at the limited level that the phenomenology of string
theories is understood, they do not necessarily yield simple gauge groups (the
essence of GUT's) or the requisite adjoint representations of the Higgs chiral
superfields which are usually necessary for the breaking of the GUT.
A typical direct product gauge group resulting as the low-energy limit of a
string theory can avoid the doublet-triplet splitting problem in GUT's.
Although one can get GUT's and adjoint (and other higher-dimensional)
representations from strings by using higher-level Kac-Moody algebras on the
string worldsheet, this reintroduces the doublet-triplet splitting problem.
Of course, one still believes that at a deep level, quark and lepton mass
matrices should be related in a simple way.}
In this context, we note that the methods for the counting and allowed
placements of
the unremovable phases in the lepton sector discussed here and in Ref.
\cite{ksl} can be combined with our analogous methods for the quark sector
discussed in Ref. \cite{ksq} in the context of these unified models of lepton
and quark masses.  If various elements of the quark and lepton mass matrices
are equal, or differ only by some real factor (because of the types of
Yukawa couplings allowed by the grand
unified gauge invariance and by various other symmetries which are imposed to
restrict these couplings), some of the invariants in the quark and lepton
sectors may coincide, leading to a reduction in the total number of phases. We
gave an example of this in Ref. \cite{ksl}.

\section{Conclusions}
\label{conclusions}

    The goal of understanding fermion masses and mixing remains one of
the most important outstanding problems in particle physics.  In this paper we
have given a detailed discussion of our methods for determining the number of
unremovable phases in a given model of leptonic mass matrices, including new
results on higher order invariants and illustrative applications to specific
models.  As ongoing and future experiments searching for evidence of neutrino
masses and lepton mixing yield new information, the construction of more
tightly constrained predictive models of fermion masses will continue.  We
believe that the general tools discussed here will be of use in these studies.

   This research was supported in part by NSF grant PHY-93-09888.
\vfill
\eject

\vfill
\eject


\begin{thebibliography}{99}

\bibitem{pdg}{Particle Data Group, Review of Particle Properties, 1992 Edition
 \prd {\bf 45}II (1992); 1994 Edition, in preparation.}

\bibitem{nurev}{A. Smirnov, review talk in the 1993 International Lepton
Photon Symposium, Cornell, Aug. 1993; hep-ph/9310368.}

\bibitem{seesaw}{M. Gell-Mann, R. Slansky, and P. Ramond, in {\it
Supergravity}, (North Holland, 1979), p. 315; T. Yanagida, in
{\it Proceedings of the Workshop on Unified Theory and Baryon Number in the
Universe} (KEK, Japan, 1979).}

\bibitem{bwlrs}{B. W. Lee and R. Shrock, \prd {\bf 16} (1977) 1444.}

\bibitem{as}{C. Albright and R. Shrock, Phys. Lett. {\bf B84} (1979) 123;
C. Albright, R. Shrock, and J. Smith, Phys. Rev. {\bf D20} (1979) 2177.}

\bibitem{cern}{WA95 Proposal (CHORUS; K. Winter, spokesman); WA96
Proposal (NOMAD; F. Vannucci, spokesman), CERN.}

\bibitem{ksl}{A. Kusenko and R. Shrock, Phys. Lett. {\bf B323} (1994) 18
(hep-ph/9311307).}

\bibitem{ksq}{A. Kusenko and R. Shrock, ITP-SB-93-58 (hep-ph/9310307);
ITP-SB-93-62 (hep-ph/9401274).}

\bibitem{yr}{I-H. Lee, J. Shigemitsu, and R. Shrock, Nucl. Phys. {\bf B330}
(1990) 225; {\bf B334} (1990) 265; R. Shrock, {\it Lattice Yukawa Models}, in
{\it Quantum Fields on the Lattice}, ed. M. Creutz (World Scientific,
Singapore, 1992).}

\bibitem{args}{J. Bagger and E. Poppitz, \prl {\bf 71} (1993) 2380
(hep-ph/9307317).}

\bibitem{albright}{C. Albright and S. Nandi, FERMILAB-PUB-93-316-T.}

\bibitem{susy}{For recent reviews, see H. Haber, SCIPP-92-33
(hep-ph/9306207);
SCIPP-93-22 (hep-ph/9308209); G. Kane et al., UM-TH-93-24 (hep-ph-9312272)
and references therein.}

\end{thebibliography}
\end{document}